\documentclass[12pt]{article}
\usepackage{fullpage}
\usepackage{setspace}
\usepackage{epsf}
\usepackage{natbib}
\usepackage{epsfig}
\usepackage{amsthm}
\usepackage{color}
\usepackage{amsmath}
\usepackage{amssymb}
\usepackage{subfigure}
\usepackage{color}

\begin{document}

\title{Classification and categorical inputs with\\
treed Gaussian process models}
\author{Tamara Broderick\\
{\tt tab@stat.berkeley.edu}\\
Department of Statistics \\
University of California, Berkeley\and
Robert B.~Gramacy\\
{\tt bobby@statslab.cam.ac.uk}\\
Statistical Laboratory\\
University of Cambridge, UK}
\date{}
\maketitle

%\doublespacing

\begin{abstract}
  Recognizing the successes of treed Gaussian process (TGP) models as
  an interpretable and thrifty model for nonparametric regression, we
  seek to extend the model to classification.  Both treed models and
  Gaussian processes (GPs) have, separately, enjoyed great success in
  application to classification problems.  An example of the former is
  Bayesian CART.  In the latter, real-valued GP output may be utilized
  for classification via latent variables, which provide
  classification rules by means of a softmax function.  We formulate a
  Bayesian model averaging scheme to combine these two models and
  describe a Monte Carlo method for sampling from the full posterior
  distribution with joint proposals for the tree topology {\em and}
  the GP parameters corresponding to latent variables at the leaves.
  We concentrate on efficient sampling of the latent variables, which
  is important to obtain good mixing in the expanded parameter space.
  The tree structure is particularly helpful for this task and also
  for developing an efficient scheme for handling categorical
  predictors, which commonly arise in classification problems.  Our
  proposed classification TGP (CTGP) methodology is illustrated on a
  collection of synthetic and real data sets.  We assess performance
  relative to existing methods and thereby show how CTGP is highly
  flexible, offers tractable inference, produces rules that are easy
  to interpret, and performs well out of sample.

  \medskip
  \noindent {\bf keywords:} treed models, Gaussian process, Bayesian
  model averaging, latent variable
\end{abstract}

\section{Introduction}
\label{sec:intro}

A Gaussian process (GP) \cite[e.g.,][]{rasmu:will:2006} is a popular
nonparametric model for regression and classification that specifies
a prior over functions.  For ease of computation, typical priors
often confine the functions to stationarity. While stationarity is a
reasonable assumption for many data sets, still many more exhibit
only local stationarity.  In the latter case, a stationary model is
inadequate, but a fully nonstationary model is undesirable as well.
Inference can be difficult due to a nonstationary model's complexity,
much of which is unnecessary to obtain a good fit.  A treed Gaussian
process (TGP) \citep{gra:lee:2008}, in contrast, can take advantage
of local trends more efficiently. It defines a treed partitioning
process on the predictor space and fits distinct, but hierarchically
related, stationary GPs to separate regions at the leaves.  The treed
form of the partition makes the model particularly interpretable. At
the same time, the partitioning results in smaller matrices for
inversion than would be required under a standard GP model and
thereby provides a nonstationary model that actually facilitates {\em
faster} inference.

Recognizing the successes of TGP for regression, we seek to extend
the model to classification. Separately, both treed models and GPs
have already been been successfully applied to classification.
Bayesian methods in the first case outline a tree prior and series of
proposals for tree manipulation in a Monte Carlo sampling framework,
leading to Bayesian CART \citep{chip:geor:mccu:1998}---extending the
classical version due to \cite{brei:1984}. In the second case lies a
method by which real-valued GP output may be utilized in a
classification context \citep{neal:1998}; for some number of classes,
the real outputs of a commensurate number of GPs ($M-1$ for $M$
classes) are taken as priors for latent variables that yield
probabilities via a softmax function.  This leads to two ways to
combine trees with GPs for classification.  The data may be
partitioned once by the tree process, and then we may associate $M-1$
GPs to each region of the partition, or instead, $M-1$ separate full
TGPs may be instantiated. We propose taking the latter route in the
interests of faster mixing. We describe a Monte Carlo sampling scheme
to summarize the posterior predictive distribution. The algorithm
traverses the full space of classification TGPs using joint proposals
for tree-topology modifications {\em and} the GP parameters at the
leaves of the tree. We explore schemes for efficiently sampling the
latent variables, which is important to obtain good mixing in the
(significantly) expanded parameter space compared to the regression
case.

The remainder of the paper is outlined as follows.  We shall first
review the GP model for regression and classification in Section
\ref{sec:gp}.  Section \ref{sec:tgp} begins with a review of the
extension of GPs to nonstationary regression models by treed
partitioning, covering the basics of inference and prediction with
particular focus on the Monte Carlo methods used to integrate over
the tree structure.  We then discuss how categorical inputs may be
dealt with efficiently in this framework. This discussion represents
a new contribution to the regression literature and, we shall argue,
is particularly relevant to our classification extensions; it is the
categorical inputs that can most clearly benefit from the
interpretation and speed features offered by treed partitioning. We
are then able to describe the TGP model for classification in detail
in Section \ref{sec:ctgp}.  In Section \ref{sec:results} we
illustrate our proposed methodology on a collection of synthetic and
real data sets. We assess performance relative to existing methods
and thereby demonstrate that this nonstationary classification model
is highly flexible, offers tractable inference, produces rules that
are easy to interpret, and (most importantly) performs well out of
sample. Finally, in Section \ref{sec:discuss} we conclude with a
short discussion.

\section{Gaussian processes for regression and classification}
\label{sec:gp}

Gaussian process (GP) models are highly flexible and nonlinear models
that have been applied in regression and classification contexts for over a
decade~\citep[e.g.,][]{neal:1997,neal:1998,rasmu:will:2006}. For
real-valued $P$-dimensional inputs, a GP is formally a prior on the
space of functions $Z: \mathbb{R}^{P} \rightarrow \mathbb{R}$ such
that the function values at any finite set of input points $x$ have a
joint Gaussian distribution~\citep{stein:1999}. A particular GP is
defined by its mean function and correlation function. The mean
function $\mu(x) = \mathbb{E}(Z(x))$ is often constant or linear in
the explanatory variable coordinates: $\mu(x) = \beta^{T} f(x)$,
where $f(x) = [1, x]$. The correlation function is defined as $K(x,x') =
\sigma^{-2} \mathbb{E}\left([Z(x) - \mu(x)][Z(x') -
\mu(x')]^{T}\right)$.

We further assume that the correlation function can be decomposed
into two components: an underlying strict correlation function
$K^{*}$ and a noise term of constant and strictly positive size $g$
that is independently and identically distributed at the predictor
points.
\begin{equation*}
    K(x^{(i)},x^{(j)}) = K^{*}(x^{(i)},x^{(j)}) + g\delta_{i,j}
\end{equation*}
Here, $\delta_{i,j}$ is the Kronecker delta, and we call $g$ the
{\em nugget}. In the model, the nugget term represents a source of
measurement error. Writing the GP as $Z(x) = \mu(x) + w(x) +
\eta(x)$, we have $\mu(x)$ as the fixed mean function, $w(x)$ as a
zero-mean Gaussian random variable with correlation $K^{*}(x)$, and
$\eta(x)$ as an i.i.d.~Gaussian noise process. Computationally, the
nugget term helps ensure that $K$ remains nonsingular, allowing the
frequent inversion of $K$ without concern for numerical instability
that might otherwise plague efficient sampling from the Bayesian
posterior as necessary for the analysis that follows.

Due to its simplicity, a popular choice for $K^{*}(x,x')$ is the
squared exponential correlation
\begin{equation*}
    K^{*}(x,x') = \exp\left\{-\frac{||x-x'||^{2}}{d}\right\},
\end{equation*}
where the strictly positive parameter $d$ describes the {\em range}
(or {\em length-scale}) of the process. I.e., $d$ governs the rate
of decay of correlation as the distance $||x-x'||$ between locations
$x$ and $x'$ increases.  The underlying (Gaussian) process that
results with this choice of $K$ is both stationary and isotropic.  It
is easily extended to incorporate a vector-valued parameter
so that correlation may decay at rates that differ
depending on direction:
\begin{equation*}
    K^{*}(x,x') = \exp\left\{-\sum_{p=1}^{P}\frac{(x_{p}-x'_{p})^{2}}{d_{p}}\right\}.
\end{equation*}
The resulting process is still stationary, but it is no longer
isotropic.  Further discussion of correlation structures for GPs can
be found in \citet{abraham:1997} or \citet{stein:1999}. These
structures include the popular Mat\'{e}rn class \citep{mate:1986},
which allows the smoothness of the process to be parameterized.  The
choice of correlation function is entirely up to the
practitioner---the methods described herein apply generally for any
$K(\cdot, \cdot)$.

The GP model, described above for the problem of regression, can be
extended to classification by introducing latent
variables~\citep{neal:1997}. In this case, the data consist of
predictors $X$ and classes $C$. Suppose there are $N$ data points and
$M$ classes. We introduce $M$ sets of latent variables
$\{Z_{m}\}_{m=1}^{M}$, one for each class. For a particular class
$m$, the generative model is a GP as before: $Z_{m} \sim
\mathcal{N}_{N}(\mu_{m}(X), K_{m}(X,X))$. The class probabilities are
now obtained from the latent variables via a softmax function
\begin{equation}
    \label{eq:softmax}
    p(C(x) = m) = \frac{\exp(-Z_{m}(x))}{\sum_{m'=1}^{M} \exp(-Z_{m'}(x))}.
\end{equation}
The generative model then specifies that the class at a point is drawn
from a single-trial multinomial distribution with these probabilities.
In practice, only $M-1$ GPs are required since we may fix the latent
variables of the final class to zero without loss of generality.

In this model, it remains to place priors, perhaps of a hierarchical
nature, on the parameters of each GP. The challenge of efficiently
sampling from the posterior of the latent variables also remains.
\citet{neal:1997} suggests sampling the hyperparameters with a
Gaussian random walk or by the hybrid Monte Carlo method.  Our choice
of hierarchical model, in contrast, allows almost entirely Gibbs
sampling moves~\citep{gra:lee:2008}.  Neal suggests sampling the
latent variables (individually) with adaptive rejection
sampling~\citep{gilks:wild:1992}.  We prefer to use
Metropolis--Hastings (MH) draws with a carefully chosen proposal
distribution and consider block-sampling the latent variables (see
Section \ref{sec:ctgp_esti}).

The GP model, for both regression and classification, features some
notable drawbacks. First, as mentioned above, the popular correlation
functions are typically stationary whereas we may be interested in
data which are at odds with the stationarity assumption.  Second, the
typical correlation functions are clumsy for binary or categorical
inputs.  Finally, the frequent inversions of a correlation matrix, an
$O(N^3)$ operation, are necessary for GP modeling, which makes these
analyses computationally expensive.

\section{Treed Gaussian processes}
\label{sec:tgp}

Partitioning the predictor space addresses all three of the potential
drawbacks mentioned above and offers a natural blocking strategy
(suggested by the data) for sampling the latent variables as required
in the classification model. In particular, after partitioning the
predictor space, we may fit different, independent models to the data
in each partition. Doing so can account for nonstationarity across the
full predictor space.  Partitions can occur on categorical inputs;
separating these inputs from the GP predictors allows them to be
treated in classical CART fashion. Finally, partitions result in
smaller local covariance matrices, which are more quickly inverted.

\subsection{TGP for regression}
\label{sec:rtgp}

CART~\citep{brei:1984} and BCART~(for Bayesian
CART)~\citep{chip:geor:mccu:1998,chip:geor:mccu:2002} for regression
essentially use the same partitioning scheme as TGP but with simpler
models fit to each partition described by the tree.  More precisely,
each branch of the tree---in any of these models---divides the
predictor space in two. Consider predictors $x \in \mathbb{R}^{P}$;
for some split dimension $p\in\{1,\dots,P\}$ and split value $v$,
points with $x_{p} \le v$ are assigned to the left branch, and points
with $x_{p}
> v$ are assigned to the right branch. Partitioning is recursive and
may occur on any input dimension $p$, so arbitrary axis-aligned
regions in the predictor space may be defined. Conditional on a treed
partition, models are fit in each of the leaf regions. In CART the
underlying models are ``constant'' in that only the mean and standard
deviation of the real-valued outputs are inferred. The underlying
CART tree is ``grown'' according to one of many decision theoretic
heuristics and may be ``pruned'' using cross-validation
methods. % and the predictive distribution defined by the constant
%models.
In BCART, these models may be either constant
\citep{chip:geor:mccu:1998} or linear \citep{chip:geor:mccu:2002} and,
by contrast with CART, the partitioning structure is determined by
Monte Carlo inference on the joint posterior of the tree and the
models used at the leaves.  In regression TGP (hereafter RTGP), the
leaf models are GPs, but otherwise the setup is identical to BCART.
Note that the constant and linear models are special cases of the GP
model. Thus RTGPs encompass BCART for regression and may proceed
according to a nearly identical Monte Carlo method, described shortly.

For implementation and inference in the RTGP model class, we take
advantage of the open source {\tt tgp} package
\citep{package:tgp} for {\sf R} available on CRAN \citep{cran:R}.
Details of the computing techniques, algorithms, default parameters,
and illustrations are provided by \cite{Gramacy:2007}.

\subsubsection{Hierarchical model}
\label{sec:rtgp_hier}

The hierarchical model for the RTGP begins with the tree prior. We
define this prior recursively following~\citet{chip:geor:mccu:1998}.
To assemble tree $\mathcal{T}$, start with a single (root) node. Form
a queue with this root node. Recursively, for each node $\eta$ in the
queue, decide to split $\eta$ with some probability
$p_{\mathrm{split}}(\mathcal{T},\eta)$. If $\eta$ splits, first choose
a splitting rule for $\eta$ according to the distribution
$p_{\mathrm{rule}}(\mathcal{T},\eta)$, and then put the new children
of $\eta$ in the queue. It remains to define the distributions
$p_{\mathrm{split}}(\mathcal{T}, \eta)$ and
$p_{\mathrm{rule}}(\mathcal{T},\eta)$. The choice
$p_{\mathrm{split}}(\mathcal{T},\eta) = \alpha ( 1 +
D_{\eta})^{-\beta}$, with $\alpha \in (0,1)$ and $\beta > 0$, includes
a base probability of splitting $\alpha$ and a penalty factor $( 1 +
D_{\eta})^{-\beta}$, which is exponential in the node depth $D_\eta$
in $\mathcal{T}$.  A simple distribution $p_{\mathrm{rule}}$ over
split-points is first uniform in the coordinates of the explanatory
variable space and then uniform over splitting values that result in
distinct data sets at the child nodes.  When non-informative priors
are used at the leaves of the tree it is sensible to have the tree
prior enforce a minimum data requirement for each region of the
partition(s).  We take the {\tt tgp} defaults in this respect, which
automatically determine an appropriate minimum as a function of the
predictor dimension $P$.

Now let $r\in\{1,\ldots,R\}$ index the $R$ non-overlapping regions
partitioned by the tree $\mathcal{T}$ formed from the above
procedure. In the regression problem, each region contains $N(r)$
points of data: $(X_{r},Z_{r})$. $F_{r}$ is defined to be an extension
to the predictor matrix with an intercept term: $F_{r} = (1,X_{r})$;
thus, $F_{r}$ has $P+1$ columns. A ``constant mean'' may be obtained
with $F_{r} = 1$; in this case, $P=0$ in Eq.~(\ref{eq:rtgp_hier})
below.  The generative model for the GP in region $r$ incorporates the
multivariate normal ($\mathcal{N}$), inverse-gamma ($\mathrm{IG}$), and Wishart
($W$) distributions as follows.
\begin{align}
Z_{r}|\beta_{r},\sigma^{2}_{r},K_{r}
    &\sim \nonumber
    \mathcal{N}_{N(r)}(F_{r}\beta_{r},\sigma^{2}_{r}K_{r}) &
\beta_{r}|\sigma^{2}_{r},\tau^{2}_{r},W,\beta_{0}
    &\sim \nonumber
    \mathcal{N}_{P+1}(\beta_{0},\sigma^{2}_{r}\tau^{2}_{r}W) \\
\sigma^{2}_{r} &  \label{eq:rtgp_hier}
\sim \mathrm{IG}(\alpha_{\sigma}/2, q_{\sigma}/2) &
    \beta_{0} & %\nonumber
    \sim \mathcal{N}_{P+1}(\mu,B) \\
\tau^{2}_{r} & \nonumber
\sim \mathrm{IG}(\alpha_{\tau}/2,q_{\tau}/2) &
    W^{-1} & \nonumber
\sim W((\rho V)^{-1}, \rho)
\end{align}
The hyperparameters $\mu, B, V, \rho, \alpha_{\sigma}, q_{\sigma},
\alpha_{\tau}, q_{\tau}$ are constant in the model, for which we take
the {\tt tgp} defaults.

\subsubsection{Estimation}
\label{sec:rtgp_esti}

Our aim in this section is to approximate the joint distribution of
the tree structure $\mathcal{T}$, the $R$ sets of GP parameters
$\{\theta_{r}\}_{r \in
  \{1,\ldots,R\}}$ in each region depicted by $\mathcal{T}$, and GP
hyperparameters $\theta_{0}$ (those variables in
Eq.~(\ref{eq:rtgp_hier}) that are not treated as constant but also
not indexed by $r$). To do so, we sample from this distribution using
Markov Chain Monte Carlo (MCMC). We sequentially draw
$\theta_{0}|\mathrm{rest}$, $\theta_{r}|\mathrm{rest}$ for each
$r=1,\dots,R$, and $\mathcal{T}|\mathrm{rest}$.  All parameters
($\theta_r$, $r=1,\dots,R$) and hyperparameters of the GPs can be
sampled with Gibbs steps, with the exception of the covariance
function parameters $\{d_{r},g_{r}\}$.  Expressions are provided by
\cite{gra:lee:2008}, and full derivations are provided by
\cite{gramacy:2005}, which are by now quite standard in the
literature so they are not reproduced here.

Monte Carlo integration over tree space, conditional on the GP
parameters $\theta_r$, $r=1,\dots,R$ is, by contrast, more involved.
Difficulties arise when we randomly draw a new tree structure from
the distribution $\mathcal{T}|\mathrm{rest}$ because it is possible
that the new tree may have more leaf nodes (or fewer) than before.
Changing the number of leaf nodes changes the dimension of
$\theta=(\theta_1,\dots,\theta_R)$, so simple MH draws are not
sufficient for $\mathcal{T}$.  Instead, reversible jump Markov Chain
Monte Carlo (RJ-MCMC) allows a principled transition between models
of different sizes~\citep{rich:gree:1997}.

However, by choosing the GP parameters at the leaf nodes of the
proposal tree carefully, the distinction between MH and RJ-MCMC can
be effectively ignored. In the case where the number of leaf nodes
does not change between the current tree state and the proposal tree,
maintaining the same collection of $\{\theta_{r}\}$ allows us to
ignore the RJ-MCMC Jacobian factor. In the case where the number of
leaf nodes increases---and the increase will always be by exactly one
leaf node---drawing the GP parameters from their priors similarly
sets the Jacobian factor in the RJ-MCMC acceptance ratio to unity;
this leaves just the usual MH acceptance ratio. \cite{gra:lee:2008}
describe proposals that facilitate moves throughout the space of
possible trees and how these moves are made reversible. The moves are
called {\em grow}, {\em prune}, {\em change}, and {\em swap}, where
the final move has a special sub-case {\em rotate} to increase the
resulting MCMC acceptance ratio. This list of four moves is largely
the same as in the BCART model~\citep{chip:geor:mccu:1998}, with the
exception of {\em rotate}. During MCMC, the moves are chosen
uniformly at random in each iteration.

\subsubsection{Prediction}
\label{sec:rtgp_pred}

From the hierarchical model in Section \ref{eq:rtgp_hier}, we can
solve for the predictive distribution of the outputs $Z$.
Conditional on the covariance structure, standard multivariate normal
theory gives that the predicted value of $Z(x\in \mathrm{region} \;
r$) is normally distributed with 
\begin{align}
 \mathbb{E}(Z(x) | \; \mbox{data}, x\in \mathrm{region}\; r ) &=
 f^\top(x) \tilde{\beta}_r + k_r(x)^\top K_r^{-1}(Z_r -
        F_r\tilde{\beta}_r), \nonumber \\
 \mbox{Var}(Z(x) | \;\mbox{data}, x\in \mathrm{region}\; r)
  &= \sigma_r^2 [\kappa(x, x) - q_r^\top(x)C_r^{-1} q_r(x)],
        \label{eq:pred} %\\
\end{align}
\vspace{-1.5cm}
\begin{align}
\mbox{where} && C_r^{-1} &= (K_r + \tau_r^2F_r W F_r^\top)^{-1} &
%&&
q_r(x) &= k_r(x) + \tau_r^2F_r W f(x) \nonumber \\
&& \kappa(x,y) &= K_r(x,y) + \tau_r^2 f^\top(x) W f(y) \nonumber
\end{align}
Here, we have $f^\top(x) = (1, x^\top)$ under the model with a linear
mean, or $f^\top(x) = 1$ under the constant mean. Further, $k_r(x)$
is an $n_r$-long vector with $k_{r,j}(x)= K_r(x, x_j)$ for all $x_j
\in \mathrm{region} \; r$, and $\tilde{\beta}_r =
\mathbb{E}(\beta|F_r,Z_r,\mathrm{rest})$ is given in closed form by
\cite{gra:lee:2008}.

Conditional on a particular tree $\mathcal{T}$, the expressions in
Eq.~(\ref{eq:pred}) betray that the predictive surface is
discontinuous across the partition boundaries of $\mathcal{T}$.
However, in the aggregate of samples collected from the joint
posterior distribution of $(\mathcal{T}, \theta)$, the mean tends to
smooth out near likely partition boundaries as the tree operations
{\em grow, prune, change}, and {\em swap} integrate over trees and
GPs with large posterior probability.  Uncertainty in the posterior
for $\mathcal{T}$ translates into higher posterior predictive
uncertainty near region boundaries.  When the data actually indicate
a non-smooth process, the treed GP retains the flexibility necessary
to model discontinuities. When the data are consistent with a
continuous process, the treed GP fits are almost indistinguishable
from continuous \citep{gra:lee:2008}.

\subsection{Categorical inputs}
\label{sec:cat_inputs}

Classical treed methods, such as CART, can cope quite naturally with
categorical, binary, and ordinal inputs.  For example, categorical
inputs can be encoded in binary, and splits can be proposed with
rules such as $x_p < 1$.  Once a split is made on a binary input, no
further process is needed, marginally, in that dimension.  Ordinal
inputs can also be coded in binary, and thus treated as categorical,
or treated as real-valued and handled in a default way.  GP
regression, however, handles such non-real-valued inputs less
naturally, unless (perhaps) a custom and non-standard form of the
covariance function is used \citep[e.g.,][]{qian:wu:wu:2008}.  When
inputs are scaled to lie in $[0,1]$, binary-valued inputs $x_p$ are
always a constant distance apart---at the largest possible distance
in the range.  When using a separable correlation function,
parameterized by length-scale parameter $d_p$, the likelihood will
increase as $d_p$ gets large if the output does not vary with $x_p$
and will tend to zero if it does (in order to best approximate a step
function correlation).  Moreover, the binary encoding of even a few
categorical variables (with several levels) would cause a
proliferation of binary inputs which would each require a unique
range parameter.  Mixing in the high dimensional parameter space
defining the GP correlation structure would consequently be poor.
Clearly, this functionality is more parsimoniously achieved by
partitioning, e.g., using a tree. However, trees with fancy
regression models at the leaves pose other problems, as discussed
below.

Rather than manipulate the GP correlation to handle categorical
inputs, the tree presents a more natural mechanism for such binary
indicators.  That is, they can be included as candidates for treed
partitioning but ignored when it comes to fitting the models at the
leaves of the tree.  They must be excluded from the GP model at the
leaves since, if ever a Boolean indicator is partitioned upon, the
design matrix (for the GP or LM) would contain a column of zeros or
ones and therefore would not be of full rank.  The benefits of
removing the Booleans from the GP model(s) go beyond producing
full-rank design matrices at the leaves of the tree. Loosely
speaking, removing the Boolean indicators from the GP part of the
treed GP gives a more parsimonious model.  The tree is able to
capture all of the dependence in the response as a function of the
indicator input, and the GP is the appropriate non-linear model for
accounting for the remaining relationship between the real-valued
inputs and outputs. Further advantages to this approach include speed
(a partitioned model gives smaller covariance matrices to invert) and
improved mixing in the Markov chain when a separable covariance
function is used since the size of the parameter space defining the
correlation structure would remain manageable.  Note that using a
non-separable covariance function in the presence of indicators would
result in a poor fit. Good range ($d$) settings for the indicators
would not necessarily coincide with good range settings for the
real-valued inputs.  Finally, the treed model allows the practitioner
to immediately ascertain whether the response is sensitive to a
particular categorical input by tallying the proportion of time the
Markov chain visited trees with splits on the corresponding binary
indicator.  A much more involved Monte Carlo technique \citep[e.g.,
following][]{SaltEtAl2008} would otherwise be required in the absence
of the tree.

If it is known that, conditional on having a treed process for the
binary inputs (encoding categories), the relationship between the
remaining real-valued inputs and the response is stationary, then we
can improve mixing in the Markov chain further by ignoring the
real-valued inputs when proposing tree operations.  In Section
\ref{sec:catreg} we shall illustrate the benefit of the treed
approach to categorical inputs in the regression context.

There is a possible drawback to allowing (only) the tree process to
govern the relationship between the binary predictors and the
response; any non-trivial treed partition would force the GP part of
the model to relate the real-valued inputs and the response separately
in distinct regions. By contrast, one global GP---with some mechanism
for directly handling binary inputs---aggregates information over the
entire predictor space. We can imagine a scenario where the
correlation structure for the real-valued inputs is globally
stationary, but the response still depends (in part) upon the setting
of a categorical input. Then partitioning upon that categorical input
could weaken the GP's ability to learn the underlying stationary
process governing the real-valued inputs.  In this case, the approach
described by \cite{qian:wu:wu:2008}, which explicitly accounts for
correlation in the response across different values of the categorical
predictors, may be preferred.  However, we shall argue in Section
\ref{sec:ctgp} that in the classification context this possibility is
less of a worry because the GP part of the model is only a prior for
the latent $Z$-values, and therefore its influence on the posterior is
rather weak compared to the class labels $C$.  Such small influences
take a back seat to the (by comparison) enormous speed and
interpretability benefits that are offered by treed partitioning on
binary inputs as illustrated empirically in Sections
\ref{sec:ctgp:synth} and \ref{sec:ctgp:real}. These benefits are not
enjoyed by the \cite{qian:wu:wu:2008} approach.

\subsection{TGP for classification}
\label{sec:ctgp}

Recall that in order to adapt the GP for classification
(Section~\ref{sec:gp}), we introduced $M-1$ sets of latent variables
$Z_{m}$ so that each predictor $X$ corresponds to $M-1$ latent
variable values. The class $C(X)$ then has a single-draw multinomial
distribution with probabilities obtained via the softmax function
(\ref{eq:softmax}) applied to these latent variables.

\subsubsection{Possible model formulations}
\label{sec:ctgp_poss}

In the TGP case, however, it is not immediately clear how these
latent variables fit into the tree structure. We can imagine at least
two possibilities. First, recall that in the RTGP model, the tree
partitions the predictor space into regions. In each region, we fit a
GP. One latent variable extension to this model would be to fit a CGP
at each leaf of the tree. Instead of a single GP, each leaf CGP
consists of $M-1$ GPs. The output of these GPs would be the $M-1$
latent variable sets in that particular region. Together, the output
over all regions would form the entire set of latent variables. Since
this model features just {\em one} tree, we call it an OTGP.

Alternatively, we might not choose to generate a CGP at each leaf of
a single tree. Rather, we could recall that the CGP amalgamates
latent variables from $M-1$ real-valued models. Instead of the $M-1$
GPs of the CGP, then, we could generate $M-1$ independent, full TGP
models. The output of the $m^{\mathrm{th}}$ TGP would be the full set
of latent variables $Z_{m}$ for the $m^{\mathrm{th}}$ class. Since
this model features multiple, independent trees, we call it an MTGP.
Note that in this latter case, the trees need not all have the same
splits. We can imagine $M-1$ perfect copies of $X$ indexed by class.
Then $(X_{m},Z_{m})$ may be partitioned differently for each $m$. In
the OTGP case, on the other hand, $(X_{m},Z_{m})$ is partitioned into
exactly the same regions across all $m$.

We choose to focus on the MTGP in the following analysis as we expect
it to exhibit better mixing than the OTGP when the number of classes
is greater than two. For a particular choice of prior and
hierarchical model, the OTGP may be thought of as an MTGP where the
trees are all constrained to split on the same variable--value pairs.
The MTGP may thus more easily model natural splits in a set
of data since each tree can be different. Imagine a data set
where class 1 occurs predominantly in region 1, class 2 occurs
predominantly in region 2, and class 3 in region 3.
As we will see in Section \ref{sec:ctgp:synth}, a single
split in each of the two MTGP trees may be sufficient to capture
these divisions. For OTGPs, however, the overarching tree in the same
case would require two splits to represent the partition. But a
greater number of splits is less likely under the given tree prior
(Section~\ref{sec:rtgp_hier}).

Moreover, the final splits in the MTGP are more interpretable since
we immediately see which class label was relevant for a given split.
In the OTGP, a split may be primarily relevant to a particular class
label or set of class labels, but it will still be applied across all
class labels.  Along similar lines, when a {\em grow} or {\em change}
move is proposed in the tree, evidence from all of the GPs of the
OTGP accumulates for or against the move. Thus it is less likely that
the acceptance probability will be high enough for such a move to
take place in the OTGP. But for the MTGP, only the latent variables
of class $m$ directly influence the proposal probability of {\em
grows} or {\em
  changes} in the $m$th tree. So it is easier for the MTGP trees to
{\em grow} and thereby take advantage of the tree structure in the
model.

Finally, as a practical consideration, the MTGP is more directly
implemented as an extension of the RTGP. The combination of tree and
latent variable sets for each class can be essentially the
same---depending on hierarchical model choice in the classification
case---as the existing RTGP formulation with the latent variables as
the outputs. Thus, the MTGP model may be coded by making minor
modifications to the the {\tt tgp} package.  Since we henceforth use
only the MTGP, we will refer to it as a CTGP (TGP for classification)
in analogy to the acronym RTGP.

\subsubsection{Hierarchical model}
\label{sec:ctgp_hier}

\begin{figure}[h]
  \centerline{
    \scalebox{1}{
      \input{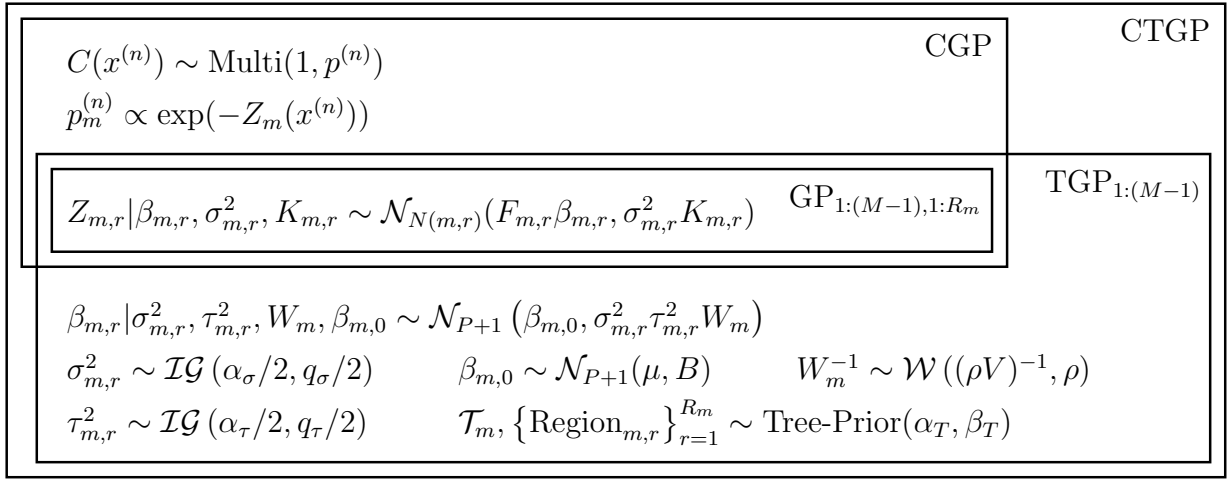}
    }
  }
  \caption{CTGP hierarchical model with GP, CGP, and TGP elements emphasized.}
  \label{fig:eg}
\end{figure}

The hierarchical model for the CTGP model is straightforward. Given
data $(X,C)$, we introduce latent variables $\{Z_{m}\}_{m=1}^{M-1}$.
There is one set for each class or, equivalently, one set for each
tree. Each of the $M-1$ trees divides the space into an independent
region set of cardinality $R_{m}$. And each tree has its own,
independent, RTGP prior where the hyperparameters, parameters, and
latent variable values---for fixed class index $m$---are generated as
in Eq.~(\ref{eq:rtgp_hier}). While this full model is identical to the
RTGP case, we prefer a constant mean for the classification task as
the linear mean incorporates extra degrees of freedom with less of a
clear benefit in modeling flexibility.  Finally, the generative model
for the class labels incorporates the latent variables from all $M-1$
trees. The softmax function appears just as in the CGP case
(\ref{eq:softmax}).  Figure \ref{fig:eg} summarizes the full model via
a plate diagram.  The devil is, of course in the details, which
follow.

\subsubsection{Estimation}
\label{sec:ctgp_esti}

To approximate the joint distribution of the $M-1$ TGPs, we sample
with MCMC as in Section~\ref{sec:rtgp_esti} to a large extent.
Sampling is accomplished by visiting each tree in turn. For the
$m^{\mathrm{th}}$ class, we sequentially draw
$\theta_{m,0}|\mathrm{rest}$, $\theta_{m,r}|\mathrm{rest}$ for each
region $r$ of $R_{m}$, $\mathcal{T}_{m}|\mathrm{rest}$, and finally
the latent variables $Z_{m,r}|\mathrm{rest}$ for each $r$.  The first
three draws are the same as for the RTGP, although they may be
modified as described below. The latent variable draw
$Z_{m,r}|\mathrm{rest}$ is the step unique to the CTGP.

One possible extension to the tree moves introduced in Section
\ref{sec:rtgp_esti}, and described in full detail
in~\citet{gra:lee:2008}, is to propose new latent variables along with
each move. In this formulation, we would accept or reject the new
partition along with the newly proposed set of latent variables and,
possibly, newly proposed GP parameters (c.f.~{\em grow}). A
difficulty with these tree moves in classification problems is that
the class data $C$ does not play a direct role; indeed, it is used
only in defining the acceptance probability of the latent variable
draws (see below). It is tempting to try to incorporate $C$ more
directly into determining the tree structure by adding latent
variable moves to {\em grow}, {\em prune}, {\em change}, and {\em
swap} (but not {\em rotate}). We could, for instance, propose $Z$
from its prior. Then the class data would appear in the posterior
probability factor of the MH acceptance ratio. The tree operations
themselves are no different whether the latent variables are proposed
or not.

However, some experimental trials with this modification suggested
that the resultant mixing in tree space is poor. Intuition seems to
corroborate this finding. Incorporating many new random variables into
a proposal will, generally, decrease the acceptance ratio for that
proposal. Certainly, when the $Z$ are all proposed simultaneously in
each leaf from their GP prior, it is unlikely that they will
predominantly arrange themselves to be in close correspondence with
the class labels. Even when a blocking mechanism is used, all of the
latent variables must be accepted or rejected together---along with
the proposed tree move.  Instead of handicapping the tree moves in
this way, we therefore separate out the latent variable proposals.
Moreover, we suspect that the tree will be most helpful for
categorical inputs, which are not dealt with in a parsimonious way by
the existing CGP model.  Indeed, the latent variables have enough flexibility
to capture the relationship between the real-valued inputs
and the class labels in most cases.

\subsubsection{Class latent variables given the class tree and
  parameters}

While we cannot sample directly from $Z_{m,r}|\mathrm{rest}$ to obtain
a Gibbs sampling draw, a particular factorization of the full
conditional for some subset of $Z_{m,r}$ suggests a reasonable
proposal distribution for MH. More specifically, the posterior is
proportional to a product of two factors: (a) the probability of the class
given the latent variables at its predictor(s)
$p(C(x_{I})|\{Z_{m,r}(x_{I})\}_{m=1}^{M-1})$ and (b) the probability of
the latent variable(s) given the current GP and other latent variables
in its region $p(Z_{m,r}(x_{I})|X_{r}, \theta,
\mathcal{T},Z_{m,r}(x_{-I}))$.
Here $I$ is an index set over the predictors $x$, and as usual $r$
labels the region within a particular class; $-I$ is the index set of
points in region $r$ of class $m$ that are not in $I$. The latter
distribution is multivariate normal, with dimension equal to the size
of the latent variable set being proposed, $|I|$. To condense
notation in Eq.~(\ref{eq:latent_distr}), let $Z_{I} = Z_{m,r}(x_{I})$
(similarly for $F$ and $\tilde{\beta}$), and let $K_{I,I'} =
K_{m,r}(x_{I},x_{I'})$. Then we have
\begin{eqnarray}
   Z_{I} &\sim& \mathcal{N}_{|I|}(\hat{z}, \hat{\sigma}^{2}) \nonumber \\
    \hat{z} &=& F_{I} \tilde{\beta}_{-I} +
    K_{I,-I}
    K^{-1}_{-I,-I}(Z_{-I}-F_{-I}\tilde{\beta}_{-I})  \label{eq:latent_distr} \\
   \hat{\sigma}^{2} &=& \sigma^{2}_{r}(\kappa_{I,I} - \kappa_{I,-I}
    \kappa_{-I,-I}^{-1} \kappa_{-I,I})  \nonumber \\
    \kappa_{I,I'} &=& k_{I,I'} + \tau^{2}_{r} F_{I} W F_{I'}^{\top} \nonumber
\end{eqnarray}
The above is essentially a condensation and slight generalization of
the predictive distribution for the real-valued output of an
RTGP~(\ref{eq:pred}). To complete the description, we again slightly
generalize the previous definitions of $V_{\tilde{\beta}}$ and
$\tilde{\beta}$~\citep{gra:lee:2008}. Here, $\tilde{\beta}_{I}$ and
$V_{I}$ are defined by their role in the distribution $\beta | X_{I},
Z_{I}, \theta, \mathcal{T} \sim \mathcal{N}_{P+1}(\tilde{\beta}_{I},
V_{\tilde{\beta},I})$. The explicit formulas are as follows.
\begin{eqnarray*}
    V_{\tilde{\beta},I}^{-1} &=& F_{I}^{\top} K_{I,I}^{-1} F_{I} + W^{-1} /
    \tau^{2}_{r} \\
    \tilde{\beta}_{I} &=& V_{\tilde{\beta},I} (F_{I}^{\top} K_{I,I}^{-1} Z_{I} 
    + W^{-1}
    \beta_{0} / \tau^{2}_{r})
\end{eqnarray*}
We sample from the latent variable distribution with MH steps. Such a
step begins with a proposal of new latents $Z'$ from the prior for $Z$
conditional on the GP parameters, i.e., following
Eq.~(\ref{eq:latent_distr}). Since the prior cancels with the proposal
probability in the acceptance ratio, the new latent variables $Z'$ are
accepted with probability equal to the likelihood ratio, where the
unprimed $Z$ represents the current latent variable values.  This
leaves:
\begin{eqnarray*}
    A &=& \frac{\exp(-Z'_{m,r}(x_{I}))}{\sum_{m'=1}^{M}
    \exp(-Z'_{m',r}(x_{I})) } \times \frac{\sum_{m'=1}^{M}
    \exp(-Z_{m',r}(x_{I}))}{\exp(-Z_{m,r}(x_{I}))}.
\end{eqnarray*}
It is not necessary to propose all of the latent variables in a region
at once. We may employ a blocking scheme whereby we propose some
subset of the latent variables in region $r$ conditioned on the rest;
we keep proposing for different (mutually exclusive) blocks until we
have iterated over all of the region's latent variables, in each step
conditioning on the accepted/rejected latents from the previously
processed blocks and those yet to be proposed (i.e., having the values
obtained in the previous iteration of MCMC).  There is a natural
trade-off in block size. Proposing $Z_{m,r}$ all at once leads to a
small acceptance ratio and poor mixing. But proposing each $z$
individually may result in only small, incremental changes. An
advantage of the treed partition is that it yields a natural blocking
scheme for updating the latent variables by making latents in a
particular region independent of those in other regions. While it may
be sensible to block further within a leaf, the automatic blocking
provided by the treed partition is a step forward from the CGP.

\subsubsection{Prediction}
\label{sec:ctgp_pred}

The most likely class label at a particular predictor value
corresponds to the smallest---or most negative---of the $M$ latent
variables obtained at that predictor; this observation follows from
the softmax generative model for the class labels given in
Eq.~(\ref{eq:softmax}).  Recall that the $M^{\mathrm{th}}$ latent is
fixed at zero.  We predict the class labels by keeping a record of the
predicted class labels at each round of the Monte Carlo run and then
take a majority vote upon completion.  We may furthermore summarize
the posterior multinomial distribution over the labels by recording
the proportion of times each class label had the lowest latent value
over the entire Markov chain, obtaining a full accounting of our
predictive uncertainty in a true and fully Bayesian fashion.

\section{Illustrations and empirical results}
\label{sec:results}

Before illustrating the CTGP algorithm and comparing it to CGP on
real and synthetic data, we shall illustrate the use of categorical
inputs in the regression context.  Two examples without categorical
inputs then follow in order to illustrate and benchmark CTGP against
CGP when the inputs are real-valued.  We conclude the section with a
real data set involving credit card applications and exhibiting both
types of inputs. One aim there will be to further highlight how an
appropriate handling of categorical inputs via trees facilitates
faster and more accurate Bayesian inference. But we also show how
this approach aids in identification of a particular categorical
feature to which the response (credit card approval) is sensitive.
This last type of output would have been difficult to elicit in the
classical CGP setup.

\subsection{Categorical inputs for regression TGP}
\label{sec:catreg}

Consider the following modification of the Friedman data
(\citealp{freid:1991}; see also Section 3.5 of
\citealp{Gramacy:2007}). Augment 10 real-valued covariates in the data
($x = \{x_1,x_2,\dots,x_{10}\}$) with one categorical indicator
$I\in\{1,2,3,4\}$ that can be encoded in binary as
\begin{align*}
1& \equiv (0,0,0,1) & 2 &\equiv (0,0,1,0) & 3 &\equiv (0,1,0,0) & 4 &\equiv
(1,0,0,0).
\end{align*}
Now let the function that describes the responses ($Z$), observed
with standard normal noise, have a mean
\begin{equation}
\mathbb{E}(Z|x, I) = \left\{ \begin{array}{cl}
    10 \sin(\pi x_1 x_2) & \mbox{if } I = 1 \\
    20(x_3 - 0.5)^2 &\mbox{if } I = 2 \\
    10x_4 + 5 x_5 &\mbox{if } I = 3 \\
    10x_1 + 5 x_2 +  20(x_3 - 0.5)^2 + 10 \sin(\pi x_4 x_5) &\mbox{if } I = 4
\label{eq:f1b}
\end{array} \right.
\end{equation}
that depends on the indicator $I$.  Notice that when $I=4$ the
original Friedman data is recovered, but with the first five inputs in
reverse order.  Irrespective of $I$, the response depends only upon
$\{x_1,\dots,x_5\}$, thus combining nonlinear, linear, and irrelevant
effects.  When $I=3$ the response is linear in $x$.  In the
experiments that follow we observe $Z(x)$ with i.i.d. $\mathcal{N}(0,1)$
noise.

Note that the encoding above precludes splits of the indicator
variable with exactly two values on each side. As has been noted
previously in the context of
CART~\citep{hastie:tibshirani:friedman:2001}, a full enumeration of
non-ordinal, categorical predictor splits is exponential in the range
of categorical predictor values. The encoding above---which
generalizes to an arbitrary number of predictor values---is more
limited; the growth in the number of splits is linear in the number of
predictor values. Observe that the full and restricted encodings
differ only for predictor ranges of size at least four. Arbitrary
groupings of predictor values are still achievable---although the
remaining values will not necessarily form a single, opposing group.

\begin{figure}[ht!]
\centering
\begin{minipage}{5cm}
\includegraphics[scale=0.6]{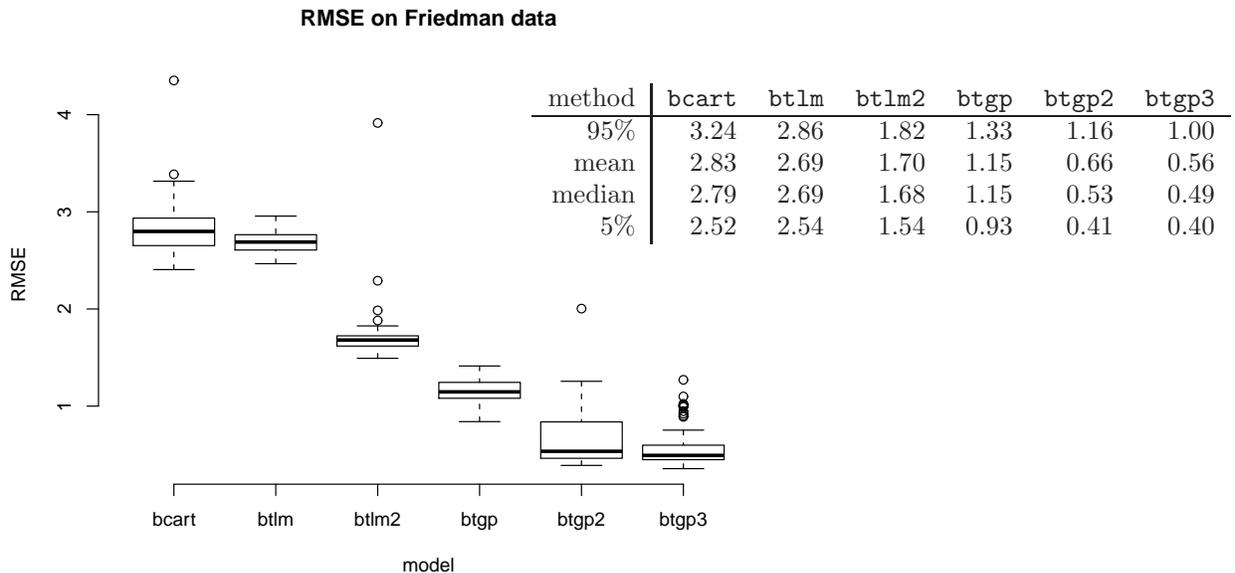}
\end{minipage}
\hfill
\begin{minipage}{9.5cm}
\footnotesize
\begin{tabular}{r|rrrrrr}
%& \multicolumn{6}{|c}{RMSE} \\
  method & {\tt bcart} & {\tt btlm} & {\tt btlm2} & {\tt btgp} & {\tt btgp2} & {\tt btgp3} \\
  \hline
  95\% & 3.24 & 2.86 & 1.82 & 1.33 & 1.16 & 1.00  \\
  mean  & 2.83 & 2.69 & 1.70 & 1.15 & 0.66 & 0.56 \\
  median & 2.79 &  2.69 & 1.68 & 1.15 & 0.53 & 0.49 \\
  5\% & 2.52 & 2.54 & 1.54 & 0.93 & 0.41 & 0.40
\end{tabular}
\vspace{3.25cm}
\end{minipage}
\caption{Summarizing root mean squared error (RMSE) obtained with
  various regression models on 100 repeated experiments with the
  adjusted Friedman data (\ref{eq:f1b}), each with a random training
  set of size 500 and a random test set of size 1000.}
\label{f:rmse}
\end{figure}

% Table \ref{t:rmse} compares summaries 
Figure \ref{f:rmse} compares boxplots and summaries of the root mean
squared error (RMSE) obtained for various regression algorithms (in
the TGP class) when trained on 500 points from (\ref{eq:f1b}) sampled
uniformly at random in $[0,1]^{10} \times \{1,2,3,4\}$ and tested on a
similarly obtained hold-out test set of size 1000.  The results and
behavior of the various methods are discussed below.

One na\"ive approach to fitting this data would be to fit a treed GP
model ignoring the categorical inputs.  But this model can only
account for the noise, giving high RMSE on a hold-out test set, and
so is not illustrated here.  Clearly, the indicators must be
included. One simple way to do so would be to posit a Bayesian CART
model ({\tt  bcart} in the Figure). In this case the indicators are
treated as indicators (as is appropriate) but in some sense so are
the real-valued inputs since only constant models are fit at the
leaves of the tree. As expected, the tree does indeed partition on
the indicators and the other inputs.

One might hope for a much better fit from a treed linear model ({\tt
  btlm}) to the data since the response is linear in some of its
inputs. Unfortunately, this is not the case. When a linear model with
an improper prior (the {\tt tgp} default) is used at the leaves of the
tree, the Boolean indicators could not be partitioned upon because
such a proposal would cause the design matrices at the two new leaves
to become rank-deficient (they would each, respectively, have a column
of all zeros or all ones). A treed GP would have the same
problem. While such models offer a more parsimonious representation of
a smoothly varying response (i.e., not piecewise constant) compared to
CART, a penalty is paid by the inability to partition; improvements in
modeling the real-valued predictors (via, for instance, the linear
model) balance against the cost of sacrificing the categorical
predictors.  Ultimately, the net result of these two effects
translates into only marginal improvements in predictive performance
over Bayesian CART.

One possible remedy is to employ a proper prior at the leaves.  But
that has its own drawbacks.  Instead, consider including the Boolean
indicators as candidates for treed partitioning, but ignoring them
when it comes to fitting the models at the leaves of the tree.  This
way, whenever the Boolean indicators are partitioned upon, the design
matrix (for the GP or LM) will not contain the corresponding Boolean
column and therefore will be of full rank.  To implement this scheme,
and similar ones to follow, we use the {\tt basemax} argument to the
{\tt b*} functions in {\tt tgp} as described by
\cite{gramacy:taddy:2009}.  Consider the treed linear model again
({\tt btlm2}) under this new treatment. In this case the MAP tree does
indeed partition on the indicators in an appropriate way---as well as
on some other real-valued inputs---and the result is the lower RMSE we
desire.

A more high-powered approach would be to treat all inputs as
real-valued by fitting a GP at the leaves of the tree ({\tt btgp}).
Binary partitions are allowed on all inputs, $X$ and $I$. As in a
direct application of the BCART linear model, treating the Boolean
indicators as real-valued is implicit in a direct application of the
TGP for regression. However, we have already seen that this
representation is inappropriate since it is known that the process
does not vary smoothly over the $0$ and $1$ settings of the four
Boolean indicators representing the categorical input $I$. Since the
design matrices would become rank-deficient if the Boolean indicators
were partitioned upon, there is no partitioning on these predictors.
As there are large covariance matrices to invert, the MCMC inference
is {\em very} slow. Still, the resulting fit (obtained with much
patience) is better than the Bayesian CART and treed LM ones, as
indicated by the RMSE.

We expect to get the best of both worlds if we allow partitioning on
the indicators but guard against rank deficient design matrices by
ignoring these Boolean indicators for the GP models at the leaves
({\tt btgp2}). Indeed this is the case, as the indicators then become
valid candidates for partitioning.

Finally it is known that, after conditioning on the indicators, the
data sampled from Friedman function are well-modeled with a GP using a
stationary covariance structure and linear mean.  We can use this
prior knowledge to build a more parsimonious model.  Moreover, the
MCMC inference for this reduced model would have lower Monte Carlo
error, since it would bypass attempts to partition on the real-valued
inputs.  These features are facilitated by the {\tt splitmin} argument
to the {\tt b*} functions, as described by
\cite{gramacy:taddy:2009}. The lower RMSE results for this model ({\tt
  btgp3}) corroborate the analysis above.

\subsection{Classification TGP on synthetic data}
\label{sec:ctgp:synth}

\subsubsection{Simple step data in 1d}
We begin with a trivial classification problem to illustrate the
difference between the CTGP and CGP models. We consider a generative
model that assigns points less than $-2/3$ to class 0, points between
$-2/3$ and $2/3$ to class 1, and points greater than $2/3$ to class 2. The
data set predictors are 60 points distributed evenly between $-2$ and
$2$, inclusive.
\begin{figure}[ht!]
\centering
\includegraphics[angle=-90,scale=0.5]{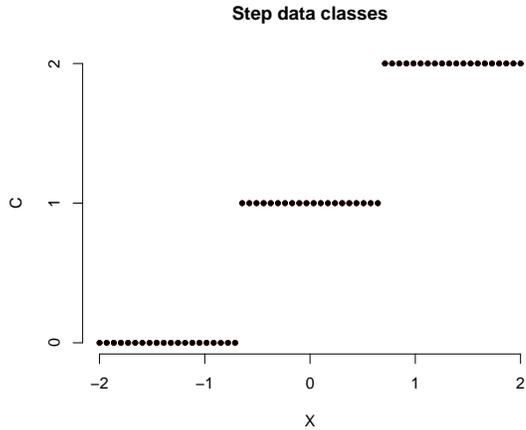}
\caption{Class labels for the step data}\label{fig:step_data}
\end{figure}
This ``step'' data is depicted in Figure \ref{fig:step_data}. Label
the points in order from $x_{0} = -2$ to $x_{59} = 2$. Clearly, we
would expect CART and BCART to form a tree with one partition between
$x_{19}$ and $x_{20}$ and another between $x_{39}$ and $x_{40}$,
perfectly dividing the classes and correctly predicting all points
outside the data set that do not fall between these dividing points.

The CGP model likewise captures the pattern of the data but does so
without the partition mechanism. In particular, running the CGP on
this data returns two sets of latent variables, one set corresponding
to class 0 and the other corresponding to class 1. Note that one
class label---here class 2---will always be represented by latent
variables that are fixed to zero. Otherwise, the model is
overdetermined (Section~\ref{sec:gp}).

\begin{figure}[htp]
\centering
\includegraphics[scale=0.5,angle=-90]{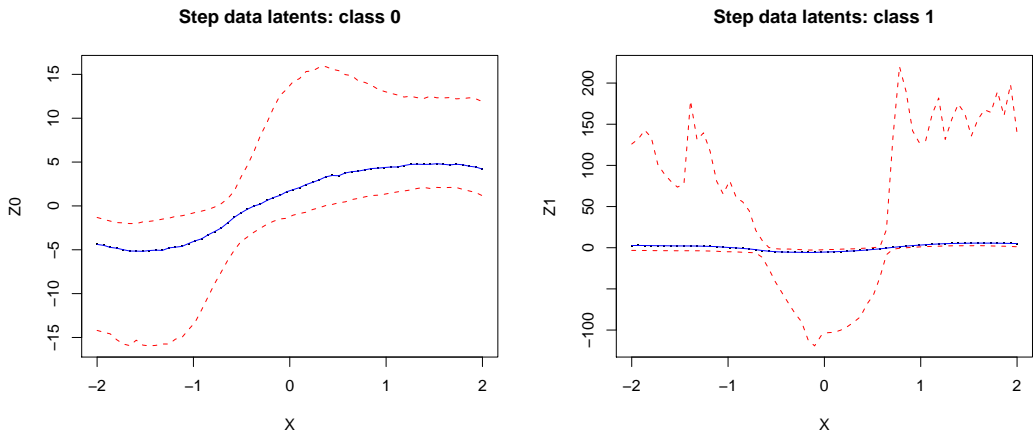}
\caption{Latent variables obtained for the step data by the CGP for
all non-trivial classes (i.e. 0 and 1). Median over MCMC rounds shown
as solid blue line; 90\% interval shown as dashed, red
lines.}\label{fig:step_cgp_latent}
\end{figure}
For the first two classes in the CGP model, the latent variable
values are depicted in Figure \ref{fig:step_cgp_latent}. As expected,
the latent variable values, which follow a Gaussian process prior,
vary smoothly across the predictor space. Since these values enter
into a softmax function to determine class probabilities in the
likelihood, smaller values roughly correspond to a higher posterior
probability for a particular class. Thus, we see that the latent
variables for class 0 mark out a negative-valued region corresponding
to the training points in class 0. There is no real distinction,
however, amongst the remaining latent variable values for this class.
Likewise, the latent variable values for class 1 mark out a
negative-valued region corresponding to the training points in class
1, and the remaining latent variable values---representing data in
other classes---are all positive and of similar modulus.

\begin{figure}[ht!]
\centering
\includegraphics[scale=0.5,angle=-90]{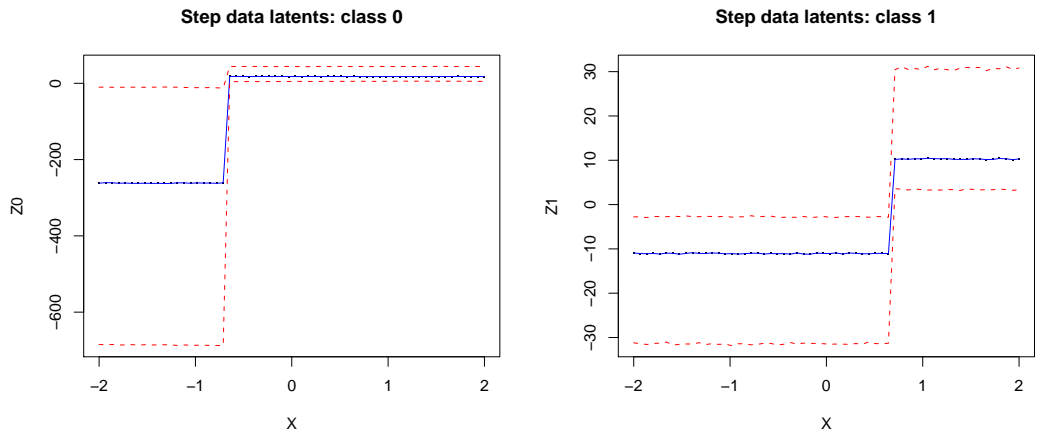}
\caption{Latent variables obtained for the step data by the CTGP for
all non-trivial classes (i.e. 0 and 1). Median over MCMC rounds shown
as solid blue line; 90\% interval shown as dashed, red
lines.}\label{fig:step_ctgp_latent}
\end{figure}

The CTGP model also captures the pattern of the data. But in this
very simple example, it reduces almost exactly to BCART. Just as for
CGP, there are two sets of latent variables---one set for class 0 and
one set for class 1 (Figure \ref{fig:step_ctgp_latent}).
Unlike the smooth variation of the CGP latent variables, the CTGP takes
advantage of the treed structure and limiting constant model. Thus,
the latent variables in class 0 exhibit a sharp divide between
$x_{19}$ and $x_{20}$. This divide sets out a negative region for
0-labeled latent variables and sets the remaining latent values at a
constant positive value. Similarly, the class 1 latent variables
exhibit a sharp divide between $x_{39}$ and $x_{40}$. In contrast to
the CGP, the divide in class 1 latent variables here separates the
data in classes 0 and 1 from the class 2 data and allows the divide
in class 0 latent variables to take care of separating the class 0
data from classes 1 and 2. This difference likely arises from the
prior penalty on larger trees. A partition that separates class 1
data from the other two classes would require a tree of height two;
the existing tree of height one accomplishes the same task with a
more parsimonious representation.

\begin{figure}[ht!]
\centering
\includegraphics[scale=0.5,angle=-90,trim=0 0 60 0]{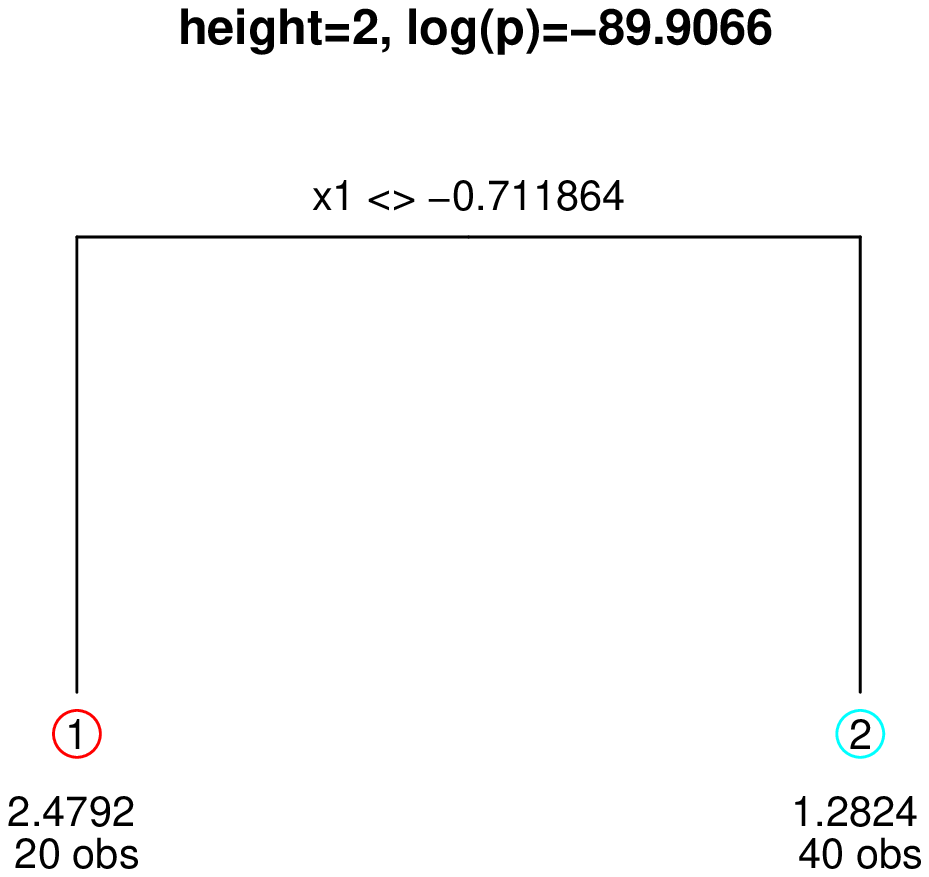}
\includegraphics[scale=0.5,angle=-90,trim=0 0 60 0]{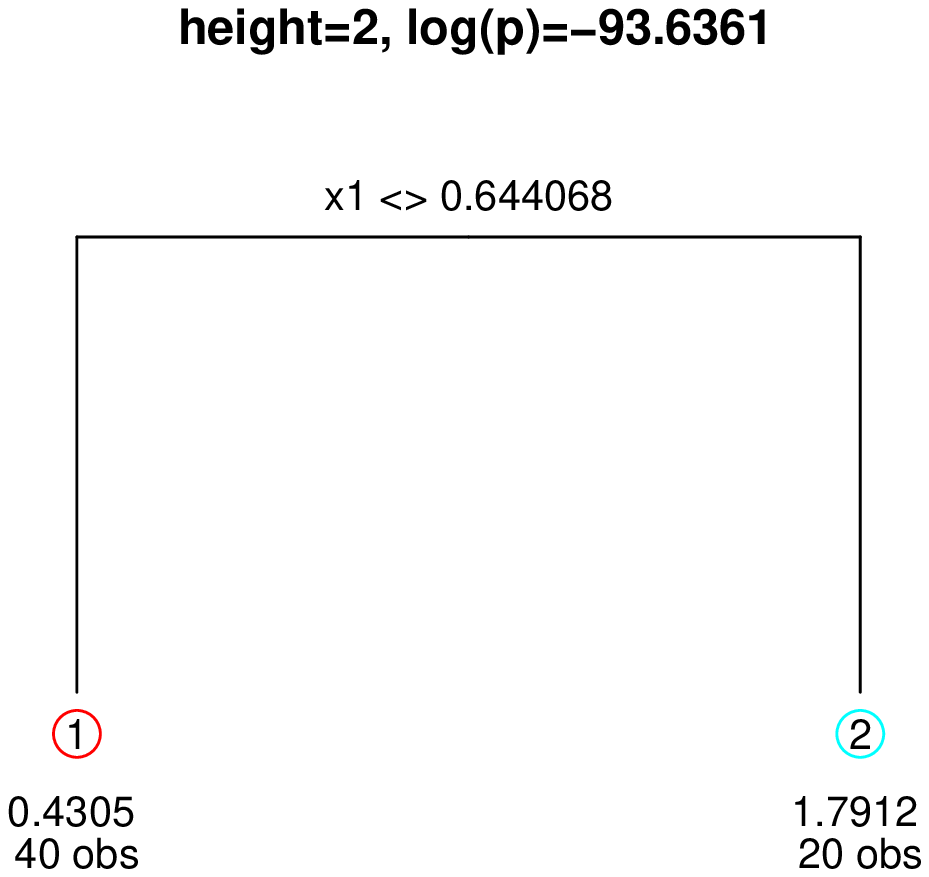}
\caption{Trees from the CTGP for the step data. {\em Left:} class 0;
{\em Right:} class 1}\label{fig:step_map1}
\end{figure}

Indeed, we can plot the maximum a posteriori trees of height greater
than one for both classes (Figure \ref{fig:step_map1}) to verify that
the divisions here are (a) caused by treed partitioning and (b) in
exactly the places we would expect given the training data. The
figure demonstrates that both of these hypotheses are indeed true.

This example illustrates that the CTGP is capable of partitioning on
continuous-valued data in practice as well as theory. Nonetheless,
this data is rather contrived and simplistic; with noisier data, it
is much less likely for real-valued partitioning to occur given our
current prior. And the partitioning does not fundamentally change the
predictive accuracy of the model. However, it does allow for a much
clearer interpretation. The treed partitioning in this example has
the same value for intuition as BCART trees whereas the CGP latent
variables, while powerful for modeling, are opaque in meaning.

Finally, we see even in this small example the running-time benefits
of the CTGP over the CGP. While the latter took $52.28$ seconds to
run, the former elapsed only $24.22$ seconds, a speedup of roughly a
factor of two.

\subsubsection{2d Exponential data}

While the CTGP easily partitions categorical inputs, as illustrated
with real data in Section \ref{sec:ctgp:real}, it can be difficult to
manufacture non-trivial synthetic classification data sets that are
obvious candidates for partitioning amongst the real-valued inputs.
In contrast, it can be quite easy to manufacture regression data
which requires a nonstationary GP model.  Here we demonstrate how a
simple Hessian calculation can convert nonstationary synthetic
regression data into class labels that are likely to benefit from
partitioning. Consider the synthetic 2d exponential data used to
illustrate RTGP by \cite{Gramacy:2007}. The input space
is $[-2,6] \times [-2,6]$, and the true response is given by
\begin{equation}
z(x) =  x_1 \exp(-x_1^2 - x_2^2). \label{e:2dtoy}
\end{equation}
In the regression example a small amount of Gaussian noise (with sd
$=0.001$) is added.  To convert the real-valued outputs to
classification labels we calculate the Hessian:
\[
H(x_1, x_2) = \begin{pmatrix}
2x_1(2x_1^2-3) \exp(-x_1^2-x_2^2) &  2x_2(2x_1^2-1) \exp(-x_1^2-x_2^2) \\
2x_2(2x_1^2-1) \exp(-x_1^2-x_2^2) & 2x_1(2x_2^2-1) \exp(-x_1^2-x_2^2)
\end{pmatrix}.
\]
Then, for a particular input $(x_1, x_2)$ we assign a class label
based on the sign of the sum of the eigenvalues of $H(x_1, x_2)$,
indicating the direction of concavity at that point.  A function like
the 2d exponential one (\ref{e:2dtoy}) whose concavity changes more
quickly in one region of the input space than in another (and is
therefore well fit by an RTGP model) will similarly have class labels
that change more quickly in one region than in another and will
similarly require a treed process for the best possible fit.
\begin{figure}[ht!]
\centering
\includegraphics[scale=0.6]{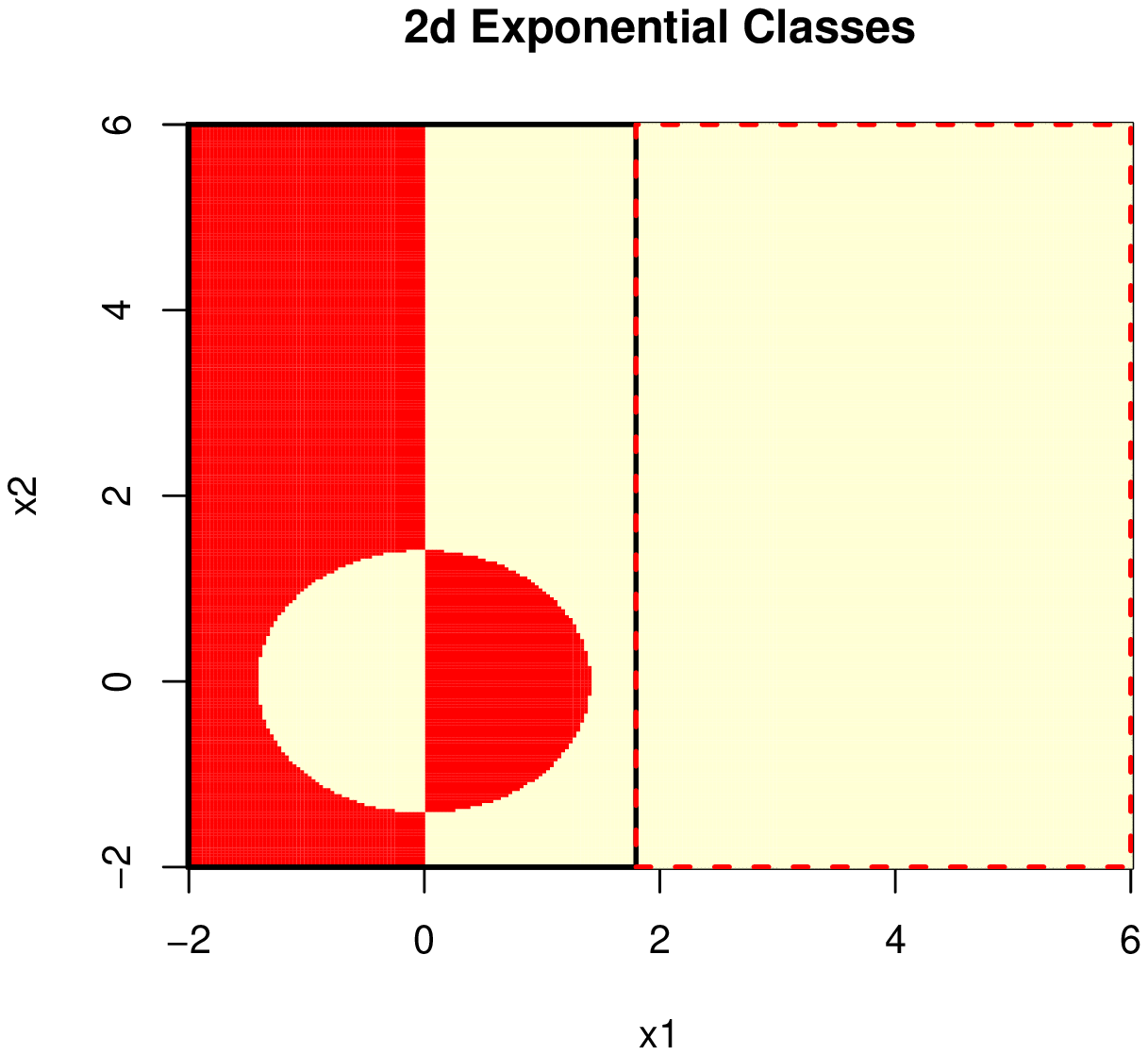} \hfill
\includegraphics[scale=0.6]{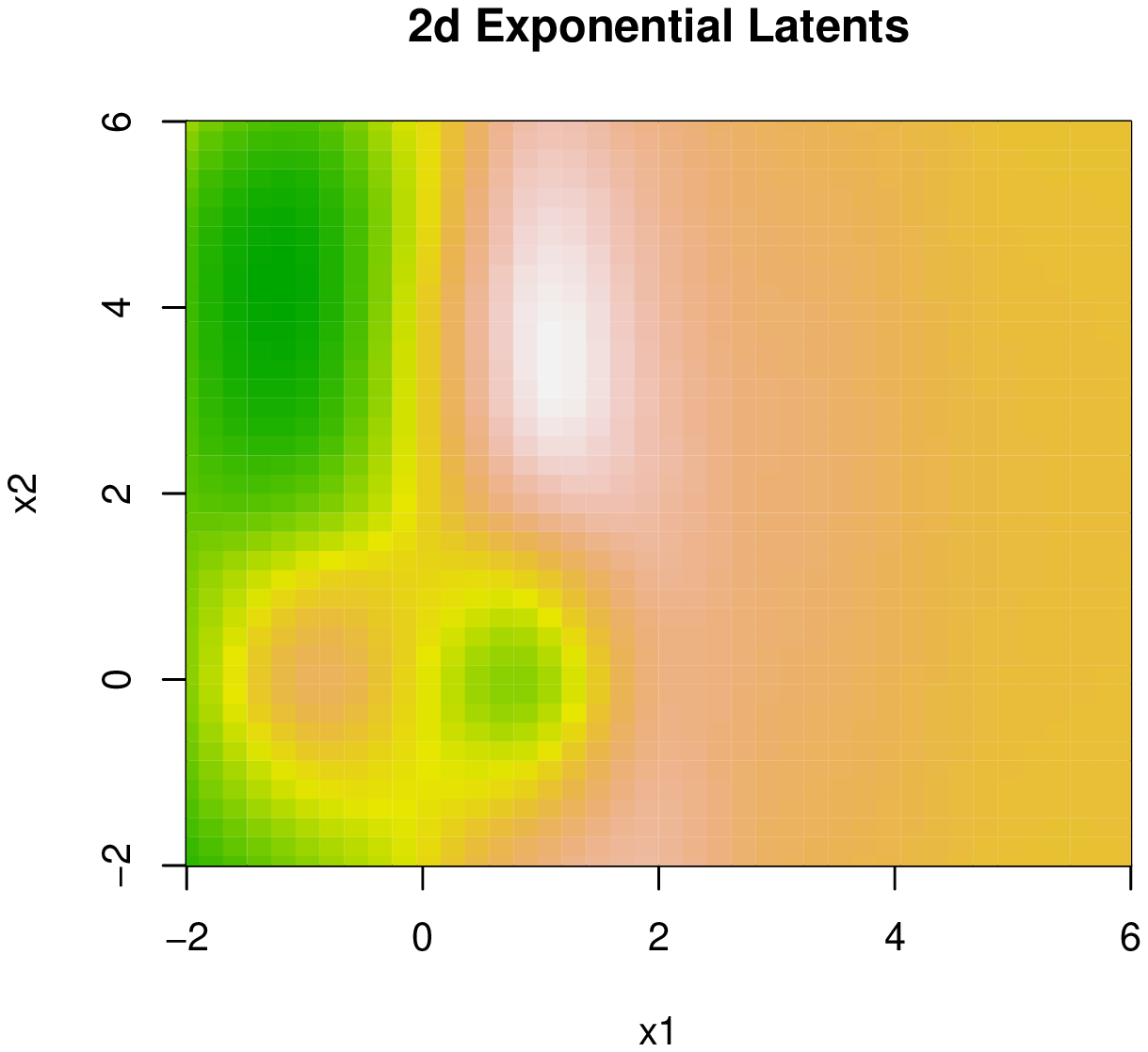}
\caption{{\em Left:} 2d exponential classification data obtained by
the Hessian eigen-method; {\em Right:} Mean latent variables
obtained by CTGP} \label{f:2dclass}
\end{figure}
Figure \ref{f:2dclass} shows the resulting class labels. Extensions
to multiple class labels (up to $m(m-1)/2$ of them) can be
accommodated by a more general treatment of (the signs of) the
components of the Hessian.

Overlaid on the plot in Figure \ref{f:2dclass} ({\em left}) is the
maximum {\em a posteriori} tree---a single split---encountered in the
trans-dimensional Markov chain sampling from the CTGP posterior.  Here
the tree is clearly separating the interesting part of the space,
where class labels actually vary, from the relatively large area where
the class is constant. As in the regression case, by partitioning, the
CTGP is able to improve upon predictive performance as well as
computational speed relative to the full CGP model.  We trained the
classifier(s) on $(X,C)$ data obtained by a maximum entropy design of
size $N=400$ subsampled from a dense grid of 10000 points and
calculated the misclassification rate on the remaining 9600 locations.
The rate was 3.3\% for CGP and 1.7\% for CTGP, showing a relative
improvement of roughly 50\%.  CTGP wins here because the relationship
between response (class labels) and predictors is clearly
nonstationary. That is, no single collection of range parameters
$\{d_1, d_2\}$ is ideal for generating the latent $Z$-values that
would best represent the class labels through the softmax.  However,
it is clear that two sets of parameters, for either side of $x_1 = 2$ as
obtained by treed partitioning, would suffice.  The speed improvements
obtained by partitioning were even more dramatic. CGP took 21.5 hours
to execute 15000 RJ-MCMC rounds whereas CTGP took 2.0 hours, an over
10-fold improvement.

\subsection{Classification TGP on real data}
\label{sec:ctgp:real}

%{\em In this experiment we only allow partitioning on the categorical
%inputs. }

The previous experiment illustrates, among other things, treed
partitions on real-valued predictors in the classification
context. However, a strength of the RTGP demonstrated in
Section~\ref{sec:catreg} is that it can offer a more natural handling
of categorical predictors compared to a GP.  With this observation in
mind, we shall compare the CTGP to the CGP (as well as neural networks
and recursive partitioning) on real data while restricting the treed
partitions in the CTGP to the categorical inputs.

For this evaluation, we considered the Credit Approval data set from
the UCI Machine Learning database~\citep{Asuncion+Newman:2007}. The
set consists of 690 instances grouped into two classes: credit card
application approval (`+') and application failure (`--'). The names
and values of the fifteen predictors for each instance are
confidential. However, aspects of these attributes relevant to our
classification task are available.  E.g., we know that six inputs are
continuous, and nine are categorical. Among the categorical
predictors, the number of distinct categories ranges from two to
fourteen. After binarization, we have a data set of 6 continuous and
41 binary predictors. The CGP treats these all as continuous
attributes. The CTGP forms GPs only over the six continuous attributes
and partitions on (and only on) the 41 binary attributes.

We also considered two other standard classification
algorithms. First, we used the neural networks (NN) algorithm as
implemented by the \texttt{nnet} function available in the
\texttt{nnet} library~\citep{ripley:2009} in {\sf R}. For each test,
we used the \texttt{tune} function in the {\sf R} library
\texttt{e1071} ~\citep{dimitriadou:2010} to simultaneously tune the
number of hidden layer units (2, 5, or 10) and the weight decay
parameter (base ten logarithm equal to -3.0, -1.5, or 0) based on the
training data. The \texttt{tune} function evaluates the classification
error for each parameter-set value via a cross-validation on the
training data and then selects the optimal parameters.  All other
settings for \texttt{nnet} parameters were kept at the default.

Second, we used recursive partitioning (RP) as implemented by the
\texttt{rpart} function ~\citep{therneau:atkinson:2010} in {\sf
  R}. Again, for each test, we used the \texttt{tune} function to
select the complexity parameter (0.01, 0.02, 0.05, 0.1, 0.2, or 0.5)
based on the training data. All other settings were kept at the
default.

\begin{table}[ht!]
\centering
\begin{tabular}{l|rr}
Method & Avg & Stddev \\ \hline
NN & 18.1 & 6.1 \\ % 1.814005e-01 6.078424e-02
RP & 15.2 & \ 3.8 \\ % 0.1523041    0.0383397
CGP & 14.6 & \ 4.0 \\ 
CTGP & 14.2 & \ 3.6
\end{tabular}
\caption{Average and standard deviation of misclassification 
  rate on the 100 credit-data folds for various methods, 
  listed in order of performance from worst to best mean.}
\label{t:real}
\end{table}

Our comparison consists of ten separate 10-fold cross-validations for
a total of 100 folds. The results are summarized in
Table~\ref{t:real}.  The CTGP offers a slight improvement over the CGP
with an average misclassification rate of 14.2\%, compared to
14.6\%. Similarly, the standard deviation of CGP misclassification
across folds was 4.0\% while the CTGP exhibited slightly smaller
variability with a standard deviation of 3.6\%. More impressive is the
speed-up offered by CTGP. The average CPU time per fold used by the
CGP method was 5.52 hours; with an average CPU time per fold of 1.62
hours, the CTGP showed a more than three-fold improvement.

%Amongst the 10 folds in the 10 repetated experiments (for a total of
%100 folds) CGP got and average missclasification rate of 0.1464012
%while CTGP got 0.1415008, a small improvement.  The variability
%amongst the folds has similar characteristics.  For CGP the variance
%was 0.001566130, while for CTGP it was 0.001325441, a small
%improvement.  More impressive is the spedups offered by CTGP.  The
%average CPU time used by the CGP method over each folds 5.52 hours,
%and 1.62 hours for CTGP, an over 3-fold improvement.

\begin{figure}
\centering
\includegraphics[scale=0.5,angle=-90, trim=0 0 60 0]{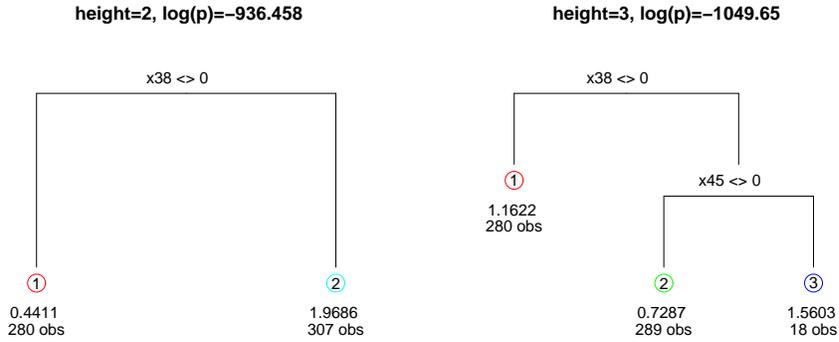}
\caption{Trees from CTGP for 2d exponential data; no height one}
\label{f:credit_trees}
\end{figure}

Finally, the interpretative aspect of the CTGP, with an appropriate
(treed) handling of the categorical inputs, is worth highlighting.
For a particular run of the algorithm on the Credit Analysis data,
the MAP trees of different heights are shown in
Figure~\ref{f:credit_trees}. These trees, and others, feature
principal splits on the $38^{\mathrm{th}}$ binary predictor, which is a
binarization of the $9^{\mathrm{th}}$ two-valued categorical predictor.
Therefore, the CTGP indicates, without
additional work, the significance of this variable in predicting the
success of a credit card application. To extract similar information
from the CGP, one would have to devise and run some additional
tests---no small feat given the running time of single CGP execution.
The figure also shows that the Markov chain visited other trees,
including one with a split on the $45^{\mathrm{th}}$ binary predictor.
However a comparison of log posteriors (shown in the Figure) and
a trace of the heights of the trees encountered in the Markov chain
(not shown) indicate that this split, and any others, had very low
posterior probability.

\section{Discussion}
\label{sec:discuss}

In this paper we have illustrated how many of the benefits of the
regression TGP model extend to classification.  The components of
TGP, i.e., treed models and GPs, have separately enjoyed a long
tenure of success in application to classification problems.  In the
case of the GP, $M-1$ processes are each used as a prior for latent
variables, which encode the classes via a softmax function.  While
the CGP is a powerful method that typically offers improvements over
simpler approaches (including treed models), drawbacks include an
implicit assumption of stationarity, clumsy handling of categorical
inputs, and slow evaluation due to repeated large matrix
decompositions.  In contrast, the treed methods are thrifty, provide
a divide-and-conquer approach to classification, and can naturally
handle categorical inputs.  Two possible ways of combining the GP
with a tree process naturally suggest themselves for, first,
partitioning up the input space and thus taking advantage of the
divide-and-conquer approach to a nonstationary prior on the latent
variables used for classification and, second, handling categorical
inputs.

We argued that it is better to work with $M-1$ separate TGP models
rather than deal with one treed model with $M-1$ GPs at each leaf.
However, a drawback of this (MTGP) approach is that, when there are
no (useful) real-valued predictors, then it will behave like BCART.
But what results will be an inefficient implementation due to all of
the latent variables involved---as the MCMC basically tries to
integrate them out.  So in this case, a direct implementation via
BCART is preferred.  However, when mixed categorical and real-valued
inputs are present, the combined tree and GP approach allows the
appropriate component of the model (tree in the case of categorical
inputs, and GP in the case of real ones) to handle the marginal
process in each input dimension.  The result is a classification
model that is speedy, interpretable, and highly accurate, combining
the strengths of GP and treed models for classification.

\subsection*{Acknowledgments} 

This work was partially supported by Engineering and Physical Sciences
Research Council Grant EP/D065704/1.  TB's research was supported by a Marshall Scholarship. We would like to thank three
anonymous referees, whose many helpful comments improved the paper.

\bibliography{tgp}
\bibliographystyle{jasa}

\end{document}